\magnification=\magstep1  
\vsize=8.5truein
\hsize=6.3truein
\baselineskip=18truept
\parskip=4truept
\vskip 18pt
\vskip 18pt
\def\today{\ifcase\month\or January\or February\or
March\or April\or May\or June\or July\or
August\or September\or October\or November\or
December\fi
\space\number\day, \number\year}
\centerline{\bf SIMULATIONS OF A LIQUID BILAYER AND HUNT FOR }
\centerline{\bf PROTRUSIONS. }
\medskip
\centerline{by }
\centerline{J. Stecki }
\vskip 20pt
\centerline{ Department III, Institute of Physical Chemistry,}
\centerline{ Polish Academy of Sciences, }
\centerline{ ul. Kasprzaka 44/52, 01-224 Warszawa, Poland}
\bigskip
\centerline{\today} 
\vskip 60pt
\centerline{Abstract. }
\vskip 20pt
New simulations are reported of a single bilayer immersed in a liquid solvent, 
using a simple extension of the model of the Max-Plack group as used in earlier
work. Fluctuation spectrum {\it vel} structure factor is dissected in
detail and the role of bulk fluctuatioms is revealed.. We propose to search 
for protrusions directly where they are, {\it i.e.} at the solvent-head boundary. 
In this context new single-point quantities and new two-point correlations are 
introduced and determined for the solvent-head pairs.
Most unusal shapes are obtained.

\vfill\eject
{\bf Introduction.}

Simulations of liquid bilayers  have been numerous and helpful in understanding 
their properties[1][2]. In one particular line of research  the structure
factor $S(q)$, readily determined in the simulation, is examined. For extended 
bilayers under lateral tension, $S$ shows the $1/q^2$ divergence as the 
transverse Fourier vector ${\bf q}$ approaches zero. It is a manifestation of 
the celebrated Goldstone mode and the capillary-wave undulations related to it. 
This capillary-wave contribution, augmented as it may be by a $1/q^4$ 
term, dies quickly with raising $q$.

Beyond that divergence,  the fluctuation spectrum 
{\it vel } structure factor is not widely nor fully understood. There is present,
 as must be, some contribution from local compression-dilatation processes 
present in all liquids as density fluctuations. Also, the polar heads in contact
 with the (polar) solvent may not make a perfect plane but may "protrude" into 
the solvent. The morphology of such "protrusions" and the dynamics of their 
formation is not known and has not been determined in any simulation. But 
otherwise it has been convincigly shown theoretically[3] that the protrusions, 
if indeed they exist, ought to modify the hydrophobic repulsion between adjacent
 bilayers in a lamellar stack. In the stack, bilayers are separated by layers 
of water - whose thickness $d$ defines various regimes of the physics of the 
lamellae. The hydrophobic repulsion has been measured in the laboratory and its 
exponential decay with $d$ is established.  

For such reasons, search for protrusions is  desirable and much needed.  
 We do that in a new way. We propose to look directly for protrusions where they 
are,  {\it i.e.} at the solvent-head boundary.

But first we examine further the structure factor $S(q)$ defined as the 
height-height correlation function. (Section 1).
As we have emphasized long time ago[4,5] it is necessary to go beyond  the
asymptotic region of the capillary wave divergence and examine a wider
range of the Fourier variable $q$. Then the nearest-neighbor peak appears[4][6].

In the search for protrusions we introduced some novel quantities, which
are shown in Section 2. But above all, instead of searching for their presence
in $S(q)$, we attempted to find them "where they are" and that was met with 
success. 

The model we use for simulations has been introduced by the Max Planck group[7]
and represents the liquid as a collection of spherical particles interacting
with various Lennard-Jones 6-12 potentials. The amphiphilic (surfactant)
molecules are made of 5 particles - beads connected by chemical
unbreakable bonds into a single chain. One terminal bead is the
hydrophilic head, the remaining 4 beads form the hydrophobic tail.
This model also goes by the name of "coarse-grained" model and is a
simplification of the so-called "Martini force field"[8]. 
Therefore we use molecular units related to the model and the potential 
used; in particular the collision diameter $\sigma$ and the depth of the 
potential minimum $\epsilon$. Thus the reduced temperature is $k_BT/\epsilon$. 

\vfill\eject
{\bf Section 1. Structure Factor and Bulk Correlations.}

The fluctuation spectrum often termed the "structure factor" $S(q)$ is an 
average height-height corrrelation; the very concept originates from a 
theoretical picture. 
In the theory, a bilayer or membrane is pictured as an infinitely thin 
sheet - a mathematical surface - embedded in three dimensions, initially  
flat "at rest" or "at equilibrium". For small deviations
from that state,  the state of the surface is fully given by a function 
$h(x,y)$.The height $h$ is measured along the $z-$axis and
 $z=z_0$ ,i.e. $h(x,y)=z_0$ is the reference flat state. 
The requirement is that 
for every $x,y$ the height $h(x,y)$ be unique,{\it i.e.} $h(x,y)$ cannot be 
a multivalued function.  Fluctuations, due to thermal motion,  
are then described by the correlation $h({\bf R}_1)h({\bf R}_2)$ .
In a translationally invariant system, the latter reduces to 
a function of the relative distance vector ${\bf R}_{12}$ and that is 
Fourier-transformed to become $S(q)$ on averaging. Here ${\bf R}$ stands for 
position vector in the $(x,y)$ plane and ${\bf q}=(q_x,q_y)$. 

Of course no membrane nor bilayer is infinitely thin; wobbling and undulations 
of the surface position $h(x,y)$ are not the only source of fluctuations. 
A fuller description of the fluctuations in any liquid is provided by the 
density-density correlation function
$$ H(1,2) \equiv \langle \rho (1) \rho (2) \rangle,   \eqno(1) $$
with  density $\rho ({\bf r})$, abbreviating
 ${\bf r}=(x_1,y_1,z_1)$ to $(1)$ etc. For nearly planar bilayers, exploiting 
translation invariance and taking the Fourier transform w.r.to the 
${\bf R}_{12}$ variable  results in  a function of three variables 
$H(z_1,z_2;q)$.
This function in inhomogeneous systems with planar interfaces or wetting layers
has been extensively investigated[9,10,11,12].

 The height fluctuation spectrum  $S(q) =\langle h_q h_{-q} \rangle $ 
is but a projection of  the density-density  correlation function: 
$$ S(q)=\int_0^L dz_1 \int_0^L dz_2 (z_1-z_0)H(z_1,z_2;q)(z_2-z_0) \eqno(2)$$
Besides providing a check on the direct computation, this route
is instructive because it demonstrates that $S(q)$ is a projection of $H$ which, 
as we know, contains bulk density fluctuations and short range correlations. 
Indeed in a homogeneous system $H$ reduces to the Fourier transform of the 
radial distribution function $g(r)$ (plus an unimportant term). 
It was obvious[4,5] that, besides the small-$q$ asymptotics of 
bending-cum-capillary waves, the traditional $S(q)$ must contain 
contributions from ordinary density fluctuations present in  all liquids. 
That it is so the nearest-neighbor peak[4,6] in $S(q)$  tells us. 
But to see it, a sufficiently wide range of $q$ must be included in the 
computation of $S$.

In simulations one deals with particles for which the microscopic density
is defined as the sum of Dirac delta-functions
$$ \rho (x,y,z) = \sum_{j=1}^N \delta (x_j -x)(y_j - y)(z_j -z) \eqno(3) $$
where $j$ is the particle index. This is converted to $\bar\rho$
obtained as the number of particles in a given box (bin) of arbitrary size;
formally it is obtained by integration over the box size.
So far we have limited the sum to particles $ j\in (1,N)$ being the beads "a"
that is the heads of the amphiphilic molecules, also distinguishing the "upper" 
from the "lower" monolayer. In principle one can construct and compute
the head-solvent (a-s), head-tail (a-t), and tail-tail (t-t) correlations but 
this has never been done so far.

It is well known that $S(q)$ diverges at $q=0^+$ showing in a spectacular way
the existence of capillary waves as $(1/q^2)$, possibly augmented by the bending 
contribution  $(1/q^4)$[13]. What happens with the increase of $q$ is not so 
well understood. In much of work only a small range of $q$ was explored;
an exception[4][6] with larger range  revealed several features. Fig.1 shows 
recent data, of the same qualitative shape. After the initial fall, $S$  goes 
through a minimum near $q^2\sim 1. $ and raises to a series of peaks, the first
 one located near $q=2\pi/\sigma$, where $\sigma$ is the collision diameter, our
unit of length. Hence the first peak is the nearest-neighbor peak, 
very much like the n.n. peak of the bulk structure factor. 

 The existence of a minimum presents a temptation to 
represent $S$ as a sum of a decreasing and an increasing term. One would be
the asymptotic divergent contribution of undulation  waves , the other would
be the contribution of bulk-like fluctuations, culminating in the 
nearest-neighbor peak.

An excellent idea is now borrowed from Reference [6] where besides $S$, the 
average with unity replacing $h$ was extracted from simulation. This was deemed
to represent the bulk-like fluctuations[6]. Formally, we obtain this function by
integrating the density-density correlation function $H$ over the $z$-variables
without the $h$ factors. There results the  generalized susceptibility, already
introduced [12] in the context of undulating liquid surfaces. First
$$ \chi(z_1,q)= \int dz_2 H(z_1,z_2;q)\eqno(4) $$
and then
$$ \chi(q) = \int dz_1 \chi(z_1,q) = \int \int dz_1 dz_2 H  \eqno(5) $$
Thus
$$ \chi(q) = \langle \sum_j \sum_m \exp[i{\bf q R}_{jm}]\rangle \eqno(6) $$
In the context of an interface in an external field, $\chi$ can be interpreted
as a susceptibility, but here it represents a certain projection of the bulk
fluctuations of the liquid bilayer.

Subtractions, i.a. of $S(q) - \chi(q)$, were also investigated [6].
In Fig.2 and 3 we show the results of such subtraction for the smoother
version of $S$ {\it i.e.} with $S$ of individual monolayers and with floating
("recentered") local reference planes. The susceptibility $\chi(q)$ is fitted to
a scaled and shifted Lorentzian about the n.n. peak. The capillary-wave asymptote
has not been subtracted hence it persists at $q\to 0$.
The difference is not nil, but even without fine-tuning the shapes show 
this is the right track - the bulk contribution is dominant  
and $\chi(q)$ is a close representation of it. One may think this
fit remarkable.

The  computation of $S$ from simulation involves a certain detail[6]
about the choice of reference plane $z=z_0$ from which $h(x,y)$ is measured.
 For a system bilayer+solvent confined to a simulation box 
 $ L\times L \times  L_z $, the obvious "absolute" coordinate 
system has the corner of the parallaepiped as its origin.
Let the position of the surface be given by $h(x,y)$. Then the flat reference
state is $h(x,y)=z_0$; the issue arises what is our choice for $z_0$? [6]. 
Briefly, it can be the middle (plane) of the bilayer $z_{mid}$ or a middle 
(average) plane of either monolayer, $\bar z_1$ or  $\bar z_2$, 
each either constant throughout the simulation run (i.e. constant in "time") or
calculated afresh at each time $t$ (i.e. "recentered" or "floating").
Explicitely then
$$ z_{mid} = {{1}\over{N}}\sum_{j\in a} z_j  \eqno(7)$$
and similarly for $z_1$ or $z_2$ with the sum limited to one monolayer. In our 
work we used the floating individual reference planes $z_1$, $z_2$ for each 
monolayer - in most cases. These choices were discussed and investigated in 
detail[6]. That the choice matters is proven by Fig.1 which shows $S(q)$ 
computed in the same run with recentered  $\bar z_1(t)$ , $\bar z_2(t)$, and 
$z_{mid}(t)$. For the first two $S$ coincide (as might be expected) but the last
shows a transition from the undulation regime to bulk fluctuation regime much 
better displayed. Such an abrupt change is therefore a universal feature as it 
was seen with two other models of intermolecular 
forces[6] and now with our version of the  "coarse-grained" model. 
The region of $q$ where $S$ raises towards the n.n. peak, corresponds in the 
bulk liquid to distances {\it larger} than the n.n. distance and therefore to 
many peaks in the r.d.f. $g(r)$. The three-dimensional Fourier transform $g(k)$ 
of the bulk liquid $g(r)$ is smooth just as $S(q)$ is in that region. 

We have also investigated the shape of a function of $q$ calculated by the 
same recipe as $S$ but for a slab of homogeneous liquid of spherical free 
particles (a) with 120000 particles at high density $\rho=0.9$ and with
periodic boundary conditions or (b) immersed in a larger cube of the same 
liquid.
In (b) boundaries of the slab were ghosts so that the slab was an open system 
with a fluctuating number of particles. For a thickness of $6$ (and $z_0$ 
constant in the middle of the slab), we obtained $S(q)$ of the same shape 
as the Percus-Yevick $ 1/(1-\rho c(q))$ curve shown in Fig.4. Also 
$S(q)$ (weighted by $(h-z_0)$) and $\chi (q)$ (not weighted cf. eq.(5-7))
differed very little in shape.See Fig.5. Such structureless background with only 
nearest-neighbor peak ought to remain if the capillary-wave contribution 
is subtracted from full $S(q)$. This exercise also showed that $\chi(q)$
was very close to these quantities obtained for a homogeneous liquid.

A technical detail is that when we compare $S(q)$ or any other quantity derived 
from $h(x,y)$ for different  reference planes,  the difference between two 
correlations
$<X(q)X^*(q)>$ (where asterisk denotes a complex conjugate) and 
$<Y(q)Y^*(q)>$, can be converted to  $<(X-Y)(X-Y)^*>$, but the mixed 
term $<X^*Y>+<XY^*>$  must be included. 
  
 As mentioned in connection with the theory of scattering off surfaces[14],
the average $<(h({\bf R}_1)-h({\bf R}_2))^2>$ is invariant with respect to the 
choice of reference plane and is equal to $ <h(R)^2> - S(R_{12})$; the 
invariance of $S(q)$ follows for $q\neq 0$. Because of imperfect orthogonality 
of the Fourier coefficients, the $S(q)$ from simulations was not invariant.  


There exists a method of computing Fourier coefficients with perfect orthogonality;
it has been employed[15] for computation of the average area of the fluctuating 
bilayer $A > L^2$ and in a calculation of the areas and curvatures[16]. An 
illustrative example can be found in the Appendix B. Given $N$ coordinates $r_j$ 
interpreted as $x_j,y_j,h(x_j,y_j)$ we choose a number $n=2m+1$ and construct 
the Fourier finite sum
$$ h({\bf R})= a_0 +\sum a_m\cos({\bf q}_m {\bf R}) + b_m\sin({\bf q}_m {\bf R})\eqno(8) $$
with the set of twodimensional $q$-vectors arbitrarily predetermined, {\it e.g.}
filling a half of a square centered on (0,0). The coefficients $a_m,b_m$ are
then determined by the least-square fit of $h$ to $N>n$ particle coordinates.
The price one pays for the perfect orthogonality is the limited number of $q$'s;
at $n=N$ the function $h(x,y)$ becomes an interpolating polynomial and this is
not what one wants. In practice $n=2m+1$ ought to be significantly smaller than
the number $N$ of points to be fitted. An example of computed curvature is given
in Appendix B.

{\bf 1A. The asymptotics of  $S(q)$}.

A quantitative description of  $S(q)$ for vanishing $q$ is
provided by the capillary-wave theory which has been formulated 
and derived in a variety of ways (see e.g.[10]) and
predicts $S_{cap}(q)= (kT/A)(1/(D+\gamma q^2)$. Here $D$
stand for the external-field contribution, $A$ is the nominal
(projected) area, $\beta=1/kT$, and
$\gamma$ is the true full macroscopic[10] surface tension.
On the other hand the very successful mesoscopic bending hamiltionian 
was combined with the above result[13].
In the absence of an external field ($D=0$) 
$$ S(q)=(kT/A)(1/( \gamma q^2 + \kappa q^4) \eqno(9) $$ 
with $\kappa$ the bending (rigidity) coefficient. 

 We  rewrite (9) as
$$ f(z)=1/(\gamma z +\kappa z^2)~~~~ z\equiv q^2 . \eqno(10) $$
Besides the  obvious pole at $z_1=0$ there  is another pole at 
$z_2=-\gamma/\kappa$. Normally $\gamma>0$, $z_2<0$ and is outside the 
range of $z$, $z\ge 0$. 
However, when we extended our simulations to the entire bilayer isotherm[4], 
we found a crossover transition to a floppy state (at areas smaller 
than that of the tensionless state), where $\gamma$ is negative.
Then the pole $z_2$  has moved to the positive part of the real axis and,
as a result, $S(q)$ diverges {\it not} at $q=0$ but at
$q^\ast =\sqrt z_2 =\sqrt{ \vert\gamma\vert/\kappa}$.

This finding[4] qualifies the standard statement about the divergence of the 
fluctuation spectrum {\it vel} structure factor $S(q)$. It is only in the extended
state with positive $\Gamma$ that $S$ diverges at $q=0^+$. 

Splitting $f(z)$ in partial fractions, we find
$$ ((z_1-z_2)f(x) = {{1}\over{(z-z_1)}} - {{1}\over{(z-z_2)}}  \eqno(11) $$
Thus the $q^{-4}$ dependence is an illusion outside the tensionless state.
At the tensionless state the two poles merge and the asymptote is a 
pure $1/z^2 = q^{-4}$.

The standard expression(9-10), proposed heuristically[13], found a support 
from  a theory[17] starting from a microscopic bending hamiltonian. 
There $\kappa$ is given; $\gamma$ appears as a derived quantity.
That calculation can be applied[2] to the  height-height 
correlation $S(q)$ and in all versions invariably the general form 
(9) results(in [2] as based on[17]). Either the intrinsic area of the 
membrane is assumed constant, or microscopic elasticity is added, either there
are three areas, four, or only two - always a form (9) is obtained, 
asymptotically at the saddle-point it must be admitted[17, 2, 16].
 There seems to be no escape from it;
unless if the theory would dispose of its basic picture of a single 
infinitely thin mathematical surface. 

The standard expression(9) does not work well as a fitting equation for the 
simulation data of S(q) at non-vanishing lateral tension $\Gamma$, even at 
apparently small $q$ where other (bulk-like) contributions are supposed 
negligible.
 If $\Gamma > 0$ also $g>0$ where we denote by $g$ the value of $\gamma$ 
obtained from a least-squares fit to $S$. Then at $q\to 0^+$, $\gamma$ wins and
$S~\sim 1/z = 1/q^2$. But adding a positive $\kappa q^4 =\kappa z^2$ term
to the denominator makes $S$ smaller, i.e. would force the curve downwards with
the second derivative negative - whereas the experimental $S$ goes gently
up above $1/z=1/q^2$. And $\kappa$ must be positive if it is to be a 
valid bending coefficient. These  defects are apparent as elucidated below.


The apparent difficulties with  (9) are  displayed in the plot of the inverse,
$1/S$ against $z=q^2$. 
The data were for $T=1.35$, $a=1.888$,$L=40$,$N=1800$ and importantly 
$\Gamma=1.16$. With such large $\Gamma$,  $\gamma$ in eq.(9) will be positive.
Therefore the curve following the data points, after starting linearly with 
positive slope should curve {\it upwards} because $\kappa>0$ and yet it did 
curve downwards.
Then we plotted $1/zS(z)$ against $z$; this is Fig.6. Here we can identify 
the very small linear part, $\gamma +z\kappa$, which agrees with eq.(9).
The point is that this range is very small.  
Having estimated the slope $\kappa$ and
 the origin $\gamma$ we go back  and add the line 
$z\gamma + z^2\kappa$, to the plot of inverse $1/S$; this is  Figure 7. 
The data agree with the theoretical 
prediction, but within a very small range of $z=q^2$.

  So the theoretical expression (9) {\it can} be used and applied, 
but its range of validity is extremely small - in the Figures essentially 
up to the first two points. Further points are "spoiled" by the non-negligible
contribution of bulk-like density fluctuations which raise $S$ and put down 
$1/S$.

In all discussion so far we have used the notion of "the average".
By the symbol $<.>$ we mean a thermal average according to a (appropriate) Gibbs
ensemble equivalently a "time" average over the simulation run; if $t$ is the 
index of a successive configuration of particles and there are $T$ such steps,
 then
$$ < X > \equiv {{1}\over{T} } \sum_t X_t \eqno(12) $$
At a given step $t$ we can introduce another average, that over particles; 
an example is an average $z$-coordinate of all particles
$$ z_{mid} = {{1}\over{N}} \sum_j z_j \eqno(13) $$
which we take as the  floating reference plane $z_{mid}(t)$. The notation $<>$ 
does not refer to such subaverages at a given time step.

The empirical  subtractions  have not shown any apparent  protrusions either 
in this work or earlier[6]. We therefore abandoned any further search for their 
signature in $S(q)$ and examined other possibilites. 
\vfill\eject  

{\bf B. Several Correlations and  Protrusions. }.

The density-density correlation $H$ but an example of two-point correlations;
 we can define  similar correlation functions for {\it any
quantity} which can be attributed to a position in space or to particle $j$.
Let it be $C$, its spatial density
$$ C(x,y,z) \equiv \sum_{j=1}^N C(j) \delta (x_j -x)(y_j - y)(z_j -z) \eqno(14) $$
 Correlations $H_{CC}$ will share several properties with the density-density 
correlation, notably its symmetry. 
We have produced several such correlation functions, mostly in our search for 
protrusions.

But before we move on to these complicated objects (functions of 3 variales)
we show first results for several one-point functions.

An  important observation is that  a protrusion is a local irregularity of the 
{\it boundary} or transition region  between the heads of a  monolayer and the 
adjacent solvent.It has not much to do with the spatial distributon of heads, 
for which it is but a detail in the wing of $\rho_a(z)$. That is why it does not show, 
apparently, in the shape of $S(q)$.  Fig.8 shows an example of  the densities
in this boundary region, for a low-temperature tensionless bilayer. Th solvent density 
$\rho_s(z)$ falls from its high value of $0.89204$ 
intersecting the left wing of the (Gaussian with very slight asymmetry) density
of heads in the "lower" monolayer. This merging of heads and solvent is a 
characteristic feature of the tensionless states; it is not always possible to 
point out to a sharp "boundary"; we  deal with smooth change.

The derivatives $d\rho_a(z)/dz = \rho^\prime_a(z)$  and $\rho^\prime_s(z)$ are more 
specific; maximum or minimum may serve as the defined position of the head-solvent boundary.
However, these quantities are averages over the plane $(x,y)$ and 
thus will not uncover any protrusion. Better candidates are their fluctuations;
off-hand one might suppose that with many protrusions forming and vanishing 
the fluctuation of $\rho^\prime_a$ ought to be larger and conversely.
The second moment can be computed easily as an average over a 
simulation run for each $z$. We do not show these results because the
protrusions did not show visibly enough.

The densitites are computed by counting the numbers of given species of particles
in boxes (bins) thus producing a histogram,  but the derivatives are not produced by
numerical differentiation but by exploiting the {\it exact} relation between 
equilibrium averages, known as the First Yvon Equation. It relates the spatial 
gradient of the first-order distribution function (density) to the average force
on that point, e.g. for the $x-$component $(d/dx) \rho(x,y,z) = < F_x(x,y,z)> $. 
Here $F_.$ denotes the force acting on the volume element $dxdydz$ about $x,y,z$. 
For our purposes this definition is transcribed to particle notation and in the 
simulation run we can make use of the forces, all calculated at each time step. Thus 
$$ d\rho_a(z)/dz =\langle ~ \sum_{j\in a} F_z(j)\delta (z-z_j) ~\rangle  \eqno(15) $$
where $F_z(j)$ is the $z-$component of the total force acting on particle $j$ and 
the brackets denote the ensemble average - in our case, average over time.
The Dirac-delta localizes the particle $j$ at $z$ and this is routinely averaged
over  bins of chosen size (0.025 in all our runs).
 Checks with numerical differentiation not only showed agreement but also a 
superiority of the force calculation. 
The fluctuation of $d\rho/dz$ is then found from 
$\langle F F \rangle -\langle F  \rangle^2 $. 

The large fluctuation  in $\rho^\prime$ was an interesting and new quantity,
but disappointing as a tool for detecting protrusions. 
Eq.(15), the first Yvon equation, suggests an examination of the 
{\it instantaneous} and/or {\it partly averaged } force and this thread
was developped; some examples are shown  below.

We have investigated several single-point 
quantities and  their histograms $P(C)dC$, e.g.
 $C = \cos (\theta )$ where $\theta$ is the angle of ${\bf F}$,  
 with  the $z$-axis, $F_z=F\cos(\theta )$; either as an average over
all pairs of chosen species, e.g. head-solvent,"a-s", or as the angle of 
the total force on particle $"j"$. For angles, the position vectors are 
equivalently used.
The wiggling of the boundary between the monolayer heads and the adjacent solvent 
ought to affect the histogram, along with the most-probable angle and the average angle.

Another possibility is to search for the coordinates $x,y,z$ of this boundary.
On the average,we obtain the density profiles and we can take the z-coordinate
of the crossing of falling $\rho_s(z)$ with raising $\rho_a(z)$ as the definition of
the position of the boundary. But the wiggling is gone on averaging. 

We have tried more sucessfully to use the midpoint of the a-s ${\bf r}_{ij}$ 
position vector of a-s pairs. 
Thus for each such pair $x_c=(x_j+x_m)/2$ with $j\in a, m\in s$ or conversely - 
with identical definitions for $y_c$ and $z_c$.
These c-points may be used to construct a surface, just as the positions of the 
heads are traditionally used to construct the surface representing the monolayer.  
But the distribution of their occurence is also of interest. The histograms 
$P(z_c)dz_c$ are shown in Fig.9 for a low temperature and 
with/without external field which
smooths the boundary and ought to damp dynamically the protrusions. The external field
(linear, derived from a double-well parabolic potential) sharpens $P$  but also 
introduces a hump on top of the expected slight asymetry. 

Having a set of positions $ x_c,y_c,z_c$ we can treat agan the $z$-coordinate as
a height and define the new $S_c$ as
$$ S_c(q) = <h_c(q)h_c(-q)>\eqno(16) $$
with
$$ h_c(q_x,q_y) = \sum_{j} (z_c(j)-z_0)\exp[i(q_x x_c(j)+q_y y_c(j)]\eqno(17) $$
and the sum over all c-points.

The resulting plot of $S_c$ vs $q$ has an entirely new shape. 
Fig.10 shows two cases: with and without external field. 
$S_c$ picks up some capillary-wave contribution, which is correct and
this is suppressed by the external field. Otherwise the field appears to 
have little effect on higher modes. Here  the nearest neighbor peak does not
appear at all. $S_c$ always shows a definite
hump in the interesting region of $q\in (1,10)$ where protrusions are 
expected to enhance the fluctuation spectrun. 

The small-gradient theory predicts[18,2] that the area increment is directly
related to $S$,
$\Delta A = A -L^2 =\sum_q A_q $ and $A_q = S(q)*q^2$. Accordingly,
we computed $S_c(q)q^2$;  the greatest increment of area precisely
falls in the region of $q$ where protrusions  are expected, as shown 
in Fig.12. For comparison, Fig.11 shows the same data before multiplication by
$q^2$, {\it i.e.} it shows the series of $S_c(q)$.
The series  range  from the floppy state to the most extended state of the bilayer 
isotherm, all at $T=1.1$. Interestingly, the hump in Fig.12  does not change 
much with the stretching/compressing of  the bilayer; the three states in the
floppy region visibly differ from the rest by an overall raise at all $q$.

The idea of protruding heads leaving temporarily their environement and venturing 
out into the neighboring solvent is  in fact closely related to the concept of the
roughness of a surface. Theory of equlibrium roughness  (see e.g.[14]) of a surface, 
in particular of a solid, is related to the picture of two points on that surface 
and on the properties when their distance greatly increases. This view is 
corroborated by our results when the correlation $S_c$ gave such promising results 
while the single-point fluctuations were not very productive.  

{\bf ~~~ Summary.}

New simulations are reported with an improved "coarse-grained" model and
force field. A  discussion is given of the bulk contributions to the 
ubiquitous fluctuation spectrum {\it vel} "structure factor". It  is shown that
in order to give any interpretation to the spectrum beyond the tiny asymptotic
region, the range of Fourier vector must include the nearest-neighbour peak 
at $q\sim 2\pi/\sigma$, at the least[4][6]. The absence of any visible signature
of the hypothetical protrusions, led to a search for such a signature at 
regions of space where these are supposed to occur, i.e. at the boundary
between solvent and heads.
 Among many possibilities which are discussed, the
new study of the head-solvent correlations appears very promising. The new
correlation function $S_c$ and the corresponding theoretical area increment
are reported and a tentative interpretation in terms of protrusions is given.

\bigskip
\bigskip

 {\bf ACKNOWLEDGEMENTS.}

 Several fruitful discussions with John F. Nagle (Pittsburgh) and 
Olle Edholm (Stockholm) are acknowledged. The interest of Professor
Robert Holyst, the present Director of the Institute of Physical Chemistry
of the Polish Academy of Sciences, is gratefully noted, and the financial
support of the Institute is also acknowledged.

\vfill\eject

 {\bf APPENDIX A }

In the Appendix we give some details about the model and simulation procedure.
 The model we use was introduced by the Max Planck group[7] as an 
extension of the very first modelling of amphiphilic molecules as dimers in 
a solvent[18]. The simulated liquid system is made of spherical particles, 
which are either solvent molecules or beads connected in  linear chains of small
 length common to all, of unbreakable (chemical) bonds, to form  amphiphilic 
molecules. One terminal bead represents the polar bead and the remaining 
4 the nonpolar hydrocarbon chain. The intermolecular energy is a sum of pair 
interactions which are cut and shifted 6-12 Lennard-Jones functions. 

In general the symmetrical matrices of parameters require 6 energies
$\epsilon_{\alpha\beta}$ and 6 collision diameters 
$\sigma_{\alpha\beta}$. Practically in all simulation work the 
collision diameters were taken the same for all particles. Often  the
energy parameters were made all equal[7,18,20,21] as well. 
Then, as one author noticed[20] the molecule was not  amphiphilic
any more.
 Most likely the sterical effects arising from impenetrability of 
beads, apparently ensured for bilayer stability.

Our model can be viewed as a simplified version of the "Martini force
field"[8], also used under the designation of a "coarse grained" model
[6]. The essential feature of an amphiphilic modecule is the 
permanent chemical bonding of two antagonistic groups - the 
hydrophilic (polar) head and the hydrophobic (non-polar) tail, 
most often a hydrocarbon tail. We denote the solvent particles as "s",
the heads as "a" and the tail beads as "t". The intramolecular bonds
are confined to the small neighbourhood of the bond distance by a 
potential well with infinite barriers. The intermolecular bonds are 
ss,sa,aa,st,at,tt and we take all $\sigma$'s equal and all $\epsilon$'s 
equal in two groups; while $\epsilon_{tt}=\epsilon_{at}=\epsilon_{st} =\epsilon$,
the other group has a stronger force field:
$\epsilon_{ss}=\epsilon_{sa}=\epsilon_{aa}=AUG*\epsilon$, with $AUG>1$.
In this way the polar beads "a" and "s" are endowed with a stronger attraction.
In all work reported here the value $AUG=2.$ was used. .
The cutoff at distance $r = 2.5\sigma$ is common to all, except that
the repulsion between {\it s-t} and {\it a-t} pairs is obtained by putting the
cutoff at $r=2^{(1/6)}$ ( and therefore the shift at $+\epsilon$).

The Molecular Dynamics simulations were done in the canonical ensemble {\it i.e.} 
at constant $T,V,N$ with Nose-Hoover thermostat, over 1-7 millions of time
steps for sizes from 49000 to 121000 particles, 1800-4500 amphiphiles.

\vfill\eject

 {\bf APPENDIX B. }
\bigskip
Here is given an illustrative example of the curvature calculation with and
without the small gradient approximation. As explained in the main text,
for 2000 heads a reasonable number of fitting coefficients would be a half
i.e. 1000 which means 500 $q$-vectors which translates to a square of 31 by 31
i.e.  for $q_x=(2\pi/L)n_x$ we can have 
$n_x=0,\pm 1,\pm2,\cdots,\pm n_{max} $ with $n_{max} \sim 15$.
That is a limitation, for values of $L$ used in simulations, nearest-neighbor
$q_{nn} = 2\pi$ is beyond reach of this procedure. Already for $n_{max}=4$ 
we have 81 by 81 matrices.
After solving for the least-square Fourier coefficients, which has to be done in quadruple
precision, the function $h(x,y)$ and its derivatives are computed as needed 
for the computation of integrals, like integral (or total)  curvature
$$ C=\int_0^L\int_0^L dx dy \sqrt{1+p_x^2+p_y^2}~{\cal H}(x,y,p_x,p_y,..)^2 \eqno(B.1) $$
where the $p$'s are the first derivatives of $h$ and the mean curvature
$\cal H$ is given by the standard expression in the Monge gauge in terms of 
first and second derivatives. Same expression with $\cal H$ omitted gives
the true area.
Small gradient approximations were reasonably close
as might be expected for computations with small $n_{max}$.  
Figure 13 shows the curvatures plotted against the areas for several $n_{max}$;
the fractal nature of the area might have been expected but the (total) mean curvature
also follows the same trend. Near $n_{max}\sim L$ the period is of the size of the bead
but we cannot reach that.

\vfill\eject

{\bf ~~~References.}
\item {[1]} J.F.Nagle and S.Tristam-Nagle, Biochemica et Biophysica\hfill\break 
            Acta  1469, 159-195 (2000).
\item {[2]} for a review of theory related to simulation (and some exotic
            correlation functions), see J. Stecki,
            Advances in Chemical Physics 144, Chapter 3 (2010).
\item {[3]} R. Lipowsky and S. Grotehans, Europhys. Lett. 23, 599 (1993).
\item {[4]} J. Stecki, J. Chem. Phys. 120, 3508 (2004).
\item {[5]} J. Stecki, cond-mat archive Dec.10,2004; J. Chem. Phys. 125, 154902 (2006).
\item {[6]} E.G.Brandt, A.R.Brown, J.N.Sachs, J.F.Nagle,and O.Edholm, Biophysical
            Journal 100, 2104 (2011) and Supplement. Where references to other work
            can be found.
\item {[7]} R. Goetz and R. Lipowsky, J. Chem. Phys. 108, 7397 (1998).
\item {[8]} S.J.Marrinck, A.H.de Vries, and A.E.Mark, J.Phys.Chem. B 108,750(2003).
\item {[9]} J. S. Rowlinson and B. Widom, Molecular Theory of Capillarity (Clarendon, 
             Oxford), 1982.
\item {[10]} R. Evans,   Adv.  Phys. 28, 143 (1979).
\item {[11]} P. Tarazona and R. Evans, Molec. Phys. 54, 1357 (1985)
\item {[12]} J.Stecki, J. Chem. Phys.  103, 9763 (1995), ibid. 108,3788(1998).
\item {[13]}  W. Helfrich and R. M. Servuss, Nuovo Cimento 3D, 137 (1984);  
           see also  Helfrich, in {\it  Les Houches, Session XLVIII, 1988,
           Liquids at Interfaces} (Elsevier, New York, 1989).
\item {[14]} S.K.Sinha, E.B.Sirota, S.Garoff, H.B.Stanley Phys.Rev. B38,2297(1988).
\item {[15]} A. Imparato, private communication.
\item {[16]} J. Stecki, unpublished.
\item {[17]} J.-B. Fourier, A. Adjari, and L. Peliti, Phys.Rev.Lett. 86, 4970(2001).
\item {[18]} A. Imparato, J. Chem. Phys. 124, 154714 (2006).
\item {[19]} B. Smit Phys. Rev. A 37, 3431 (1988)
\item {[20]} A. Imparato, J. C. Shilcock, and R. Lipowsky, Eur. Phys. J. E. 11, 21 (2003).
\item {[21]} J. Stecki, J. Chem. Phys. Comm.122, 111102 (2005).

\vfill\eject 

{\bf ~~~FIGURE CAPTIONS}
\bigskip

    Caption to Fig.1

 The $S(q)=<h.h>$ correlations for $T=1.35~ L_x=61.25~ N\sim 4000$,$\rho=0.89$, 
total 121000 particles, tail length 8. $S$ for all heads   shown
with plus signs (reference plane as recentered average $z_{mid}(t)$). 
 Otherwise  either monolayer (crosses) with
 individual recentered $\bar z_1,\bar z_2$.
The const.$/q^4$ asymptote is shown with line. The remarkable break near
$q^2=0.1$ is also present in individual monolayer $S$ provided
the reference plane is $z_{mid}$,  constant or recentered.
\medskip

   Caption to Fig.2 

The generalized susceptibility $\chi (q)$ due to bulk-like 
fluctuations is fitted to a scaled and shifted Lorentzian 
$f(q)=f_0+a/((q-q_0)^2 +b^2)$ and subtracted from the full $S(q)= <h.h>$.
Even w/o fine-tuning,  this fit proves the dominant role of of bulk 
fluctuations in $S(q)= <h.h>$. Bare $S$ - plus signs, $\chi$ - crosses,
$f$ -  line, and the difference - boxes. 
\medskip

 Caption to Fig. 3 

The subtraction of generalized susceptibility $\chi (q)$ from $S(q)=<h.h>$.
Plotted against $q$: $S(q)$  ( floating reference
plane for one monolayer) - with plus signs, $\chi$ - crosses, fit to $\chi$ by 
a Lorentzian -  line, the difference - boxes. Clearly the bulk 
contribution to $S$ follows the shape of $\chi$. Arbitrary scale, forcing 
equality of n.n.peaks.  Same data as Figs.1 and 2.
\medskip

   Caption to Fig.4

Correlation $\chi(q) =<\exp[ikR]>$ for the bilayer shown with plus signs and
for the slab immersed in LJ liquid - with boxes, plotted against $q$. 
 The identical shapes show that bulk-like fluctuations produce $\chi$ in the 
bilayer. Bulk Percus-Yevick $S=1+\rho h =1/(1-\rho c)$ is plotted with the line.
Density $\rho=0.9$.
\vfill\eject

    Caption to Fig.5 

Fluctuation spectra for a homogeneous slab of $N=$
120000 L-J particles at $T=1.1$ density $\rho=0.89$.
The reference plane  is at $z_0=L_z/2$. 
$\chi =<\rho\rho(q)>$ - plus signs,$s=<h.h>$ - boxes,
$s\prime=<h.h>$ - diamonds. $s$ for all N particles,
$s\prime$ - open system  within $L_z/4, 3L_z/4$. The corresponding 
quantities for $T=5$ are shown with stars and differ surprisingly 
little from $T=1.1$ data.
The ordinate is $n_x^2+n_y^2$ proportional to $q^2$.
 The n.n.peak is damped by the logarithmic scale. 
\medskip

    Caption to Fig.6 

Inverses $1/zS(z)$ plotted against $z=q^2$ for bilayer (boxes) and 
two monolayers (plus signs and stars) with common reference plane 
$z_0=z_{mid}$ recentered and monolayers with individual reference planes
recentered (circles). The straight line is a best fit to all, of 
$g+\kappa z$, at $z\to 0$. Data are for $T=1.35,L_x=40.,\rho=0.89,N=1800,
a=1.88889$.
\medskip

    Caption to Fig.7 

Inverses $1/S(z)$ plotted against $z=q^2$, with labels as in Fig.5,
with the parabola $g*z + \kappa *z^2$ obtained from the straight line
of Fig.5. Extremely small range of validity of Eq.(9) is apparent. But
valid it is.
\medskip

    Caption to Fig.8 

Densities $\rho(z)$ and their derivatives $\rho^\prime (z)$ 
in the lower monolayer,
shown as follows: solvent - full line falls as error function,
heads - full line and shaded vertically.
$\rho_s^\prime(z)$  line with small boxes, very accurately a gaussian. 
$\rho_a^\prime (z)$ - full line (black) with positive
and negative values. At this tensionless state at T=1.1 the boundary 
between heads and solvent is fuzzy and broad.
\medskip

  Caption to Fig.9  

Histogram of the $z$-coordinate of $r_c$ (see text) plotted
against $z$ for the lower monolayer; solvent is to the left (lower $z$) and
heads to the right, tails largely outside the Figure. No external field -
plus signs, with a linear external field - stars; with an external field also
supporting an artificial protrusion - circles. For comparison the density 
profile of heads (scaled by an arbitrary factor) is shown with small boxes 
and shaded by vertical lines.  
\vfill\eject

   Caption to Fig. 10   

The new correlation $S_c=<h_c.h_c>$ for one monolayer with floating local
reference plane with and without external field, plotted against $q^2$.
$T=1.1$, 1800 amphiphiles,near tensionless state.
 A weak external field from a double well parabolic 
potential damps capillary waves. See text.
\medskip

    Caption to Fig.11   

The $S_c=<h_c.h_c> $ correlation with $h_c$ the (Fourier component of) 
$z$-coordinate of $r_c$ plotted against $x=q^2$.The series includes floppy
states and most extended states near the breajing point.For details  
see  Caption to Fig.12
\medskip

    Caption to Fig.12

The $<h_c.h_c> $ correlation with $h_c$ the (Fourier component of) the 
$z$-coordinate of $r_c$ (see text) multiplied by $q^2$ and plotted against
$z=q^2$.In the small-gradient theory the quantity $zS(z)$ is the $q$-component 
of the area increment $dA_q$.The successive curves are (in this order) for 
$L_x=L_y$ equal to 100. (label"1"), 97., 73.5(label"3"), 64.0 ("4"), 62.5 ("5"),
61.25 ("6") 61.0 ("7") 60. ("8"). and 59.("9"). 
The last three belong to the floppy region. 
An enhancement in the region $1\le q^2 \le 10$ can be seen, largely
independent of the area per head. $T=1.35$, 121000 particles, 2000 heads 
in this monolayer.The area per head ranges from 1.7826 to 5.035.
\medskip

     Caption to Fig.13  (Appendix B)

For successively increasing numbers of Fourier coefficients (M=12,40,112,364)
total (mean) curvature and area increment (over $L^2$)  computed
from the same data on 4000 heads by the method of Appendix B. Here $\Delta A$
is plotted against total $\cal H$. There is no saturation in sight. Each point 
in the cluster of data points results from one configuration of heads,taken
succesively from the same simulation run, at T=1.35,density 0.89 $L=39.$
2000 heads in one monolayer.
\medskip

\vfill\eject\end
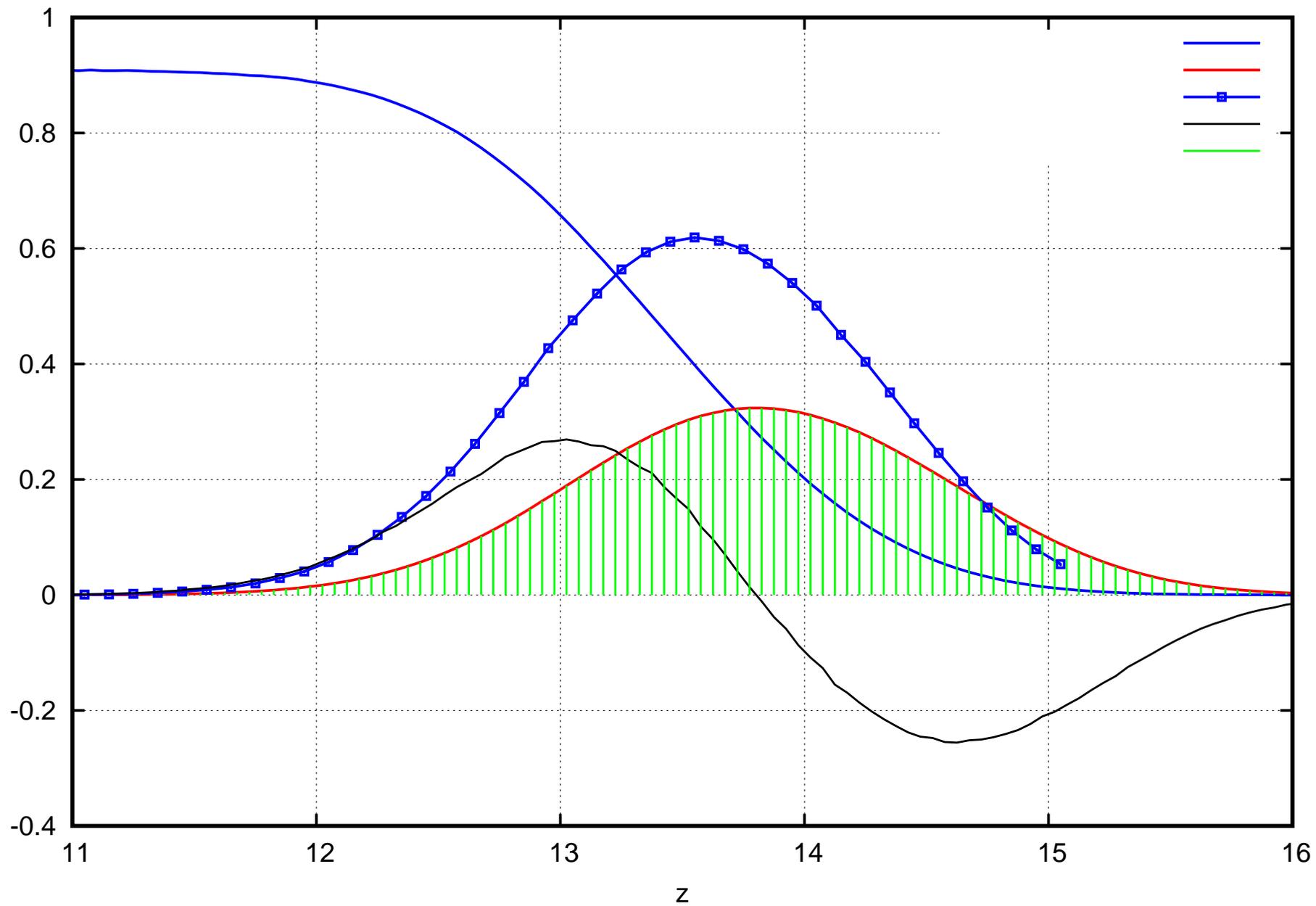
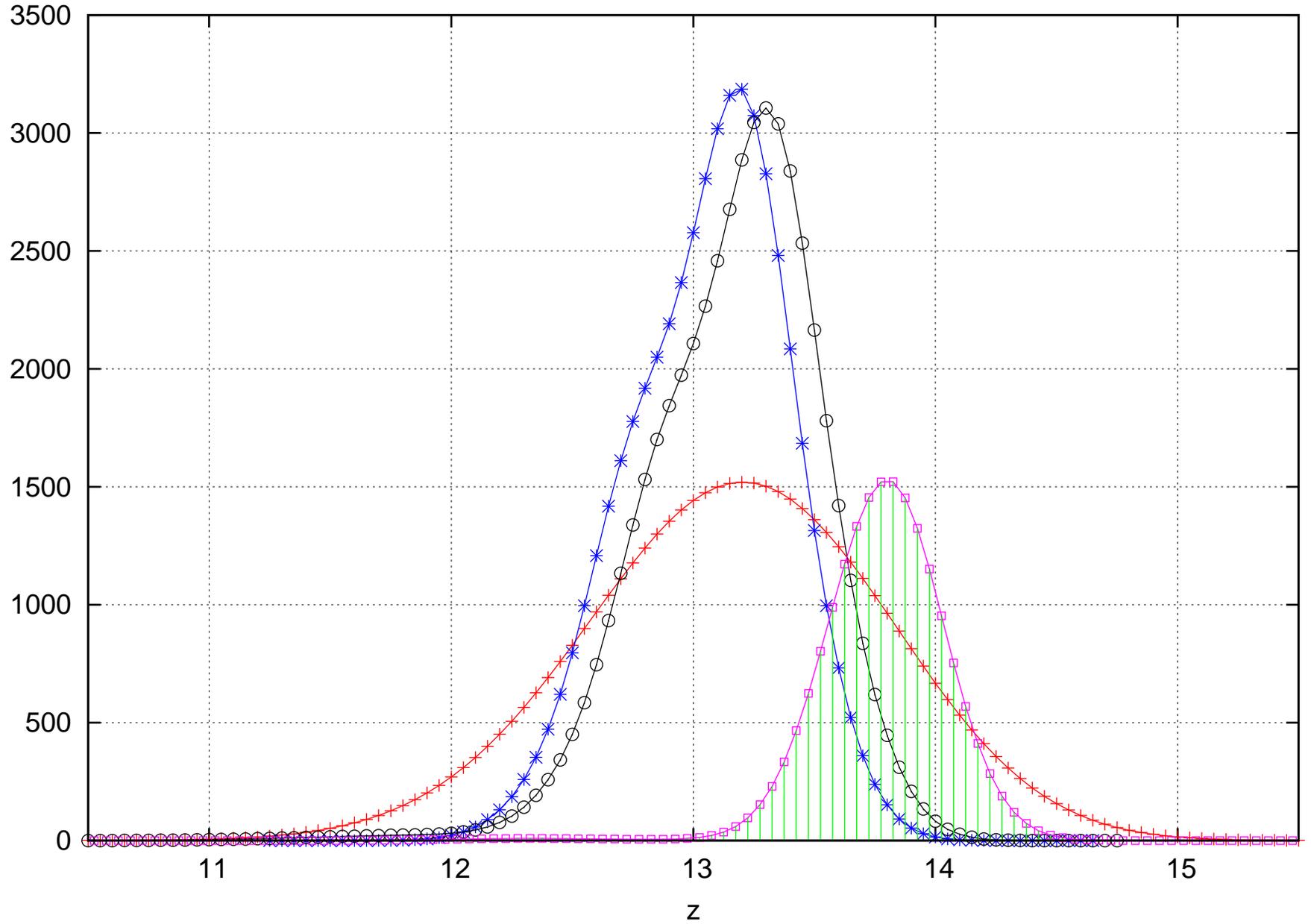
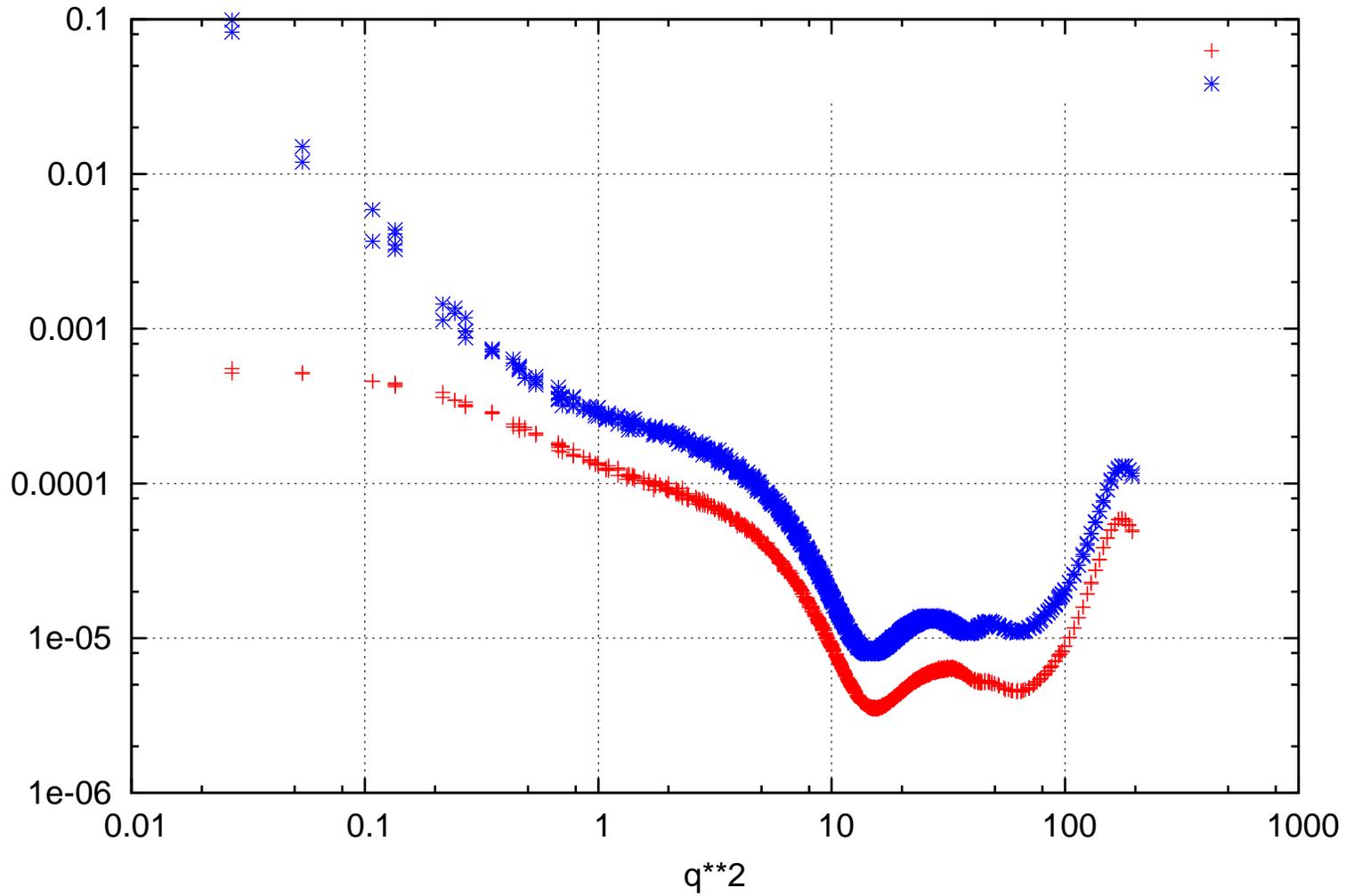
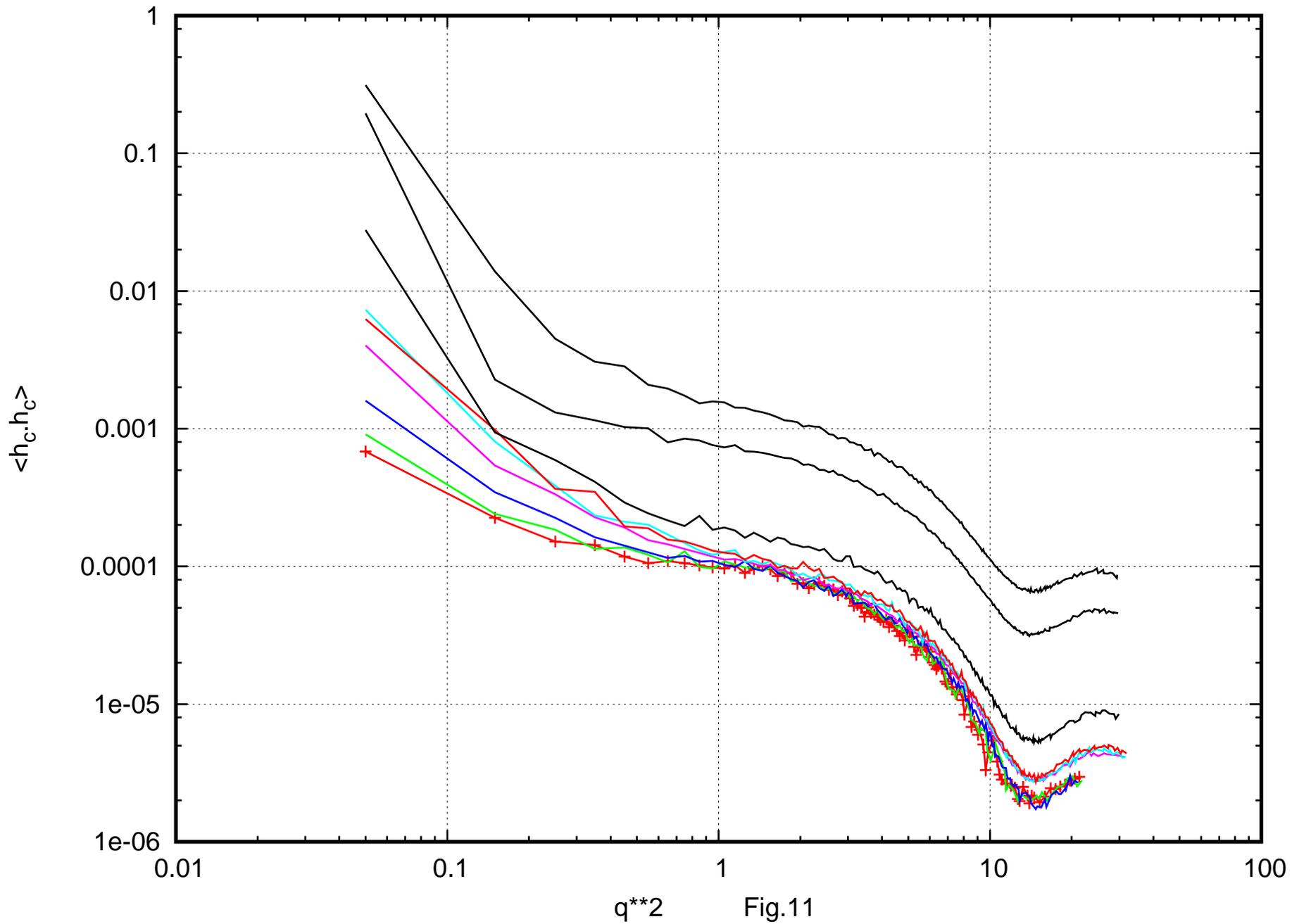
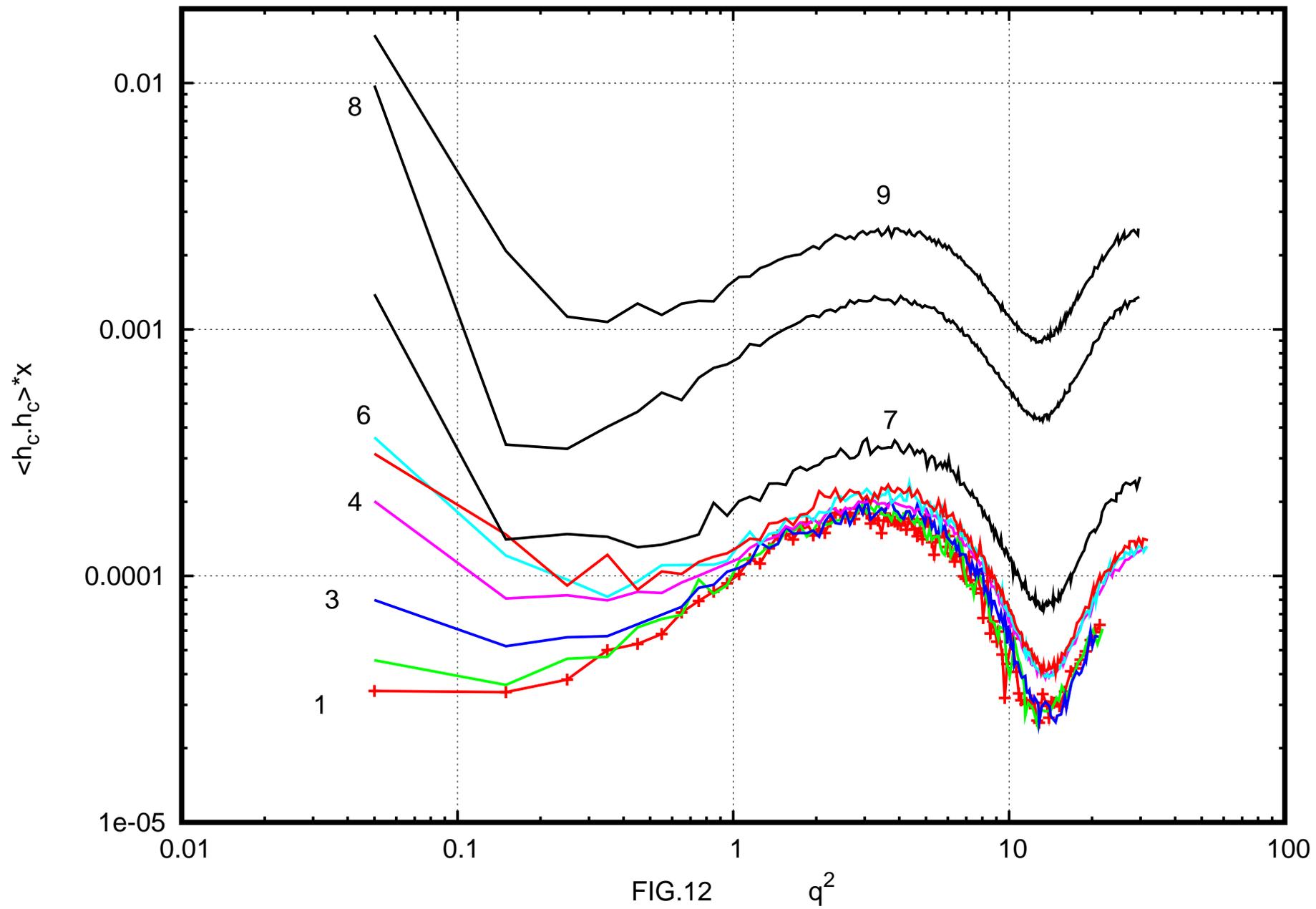
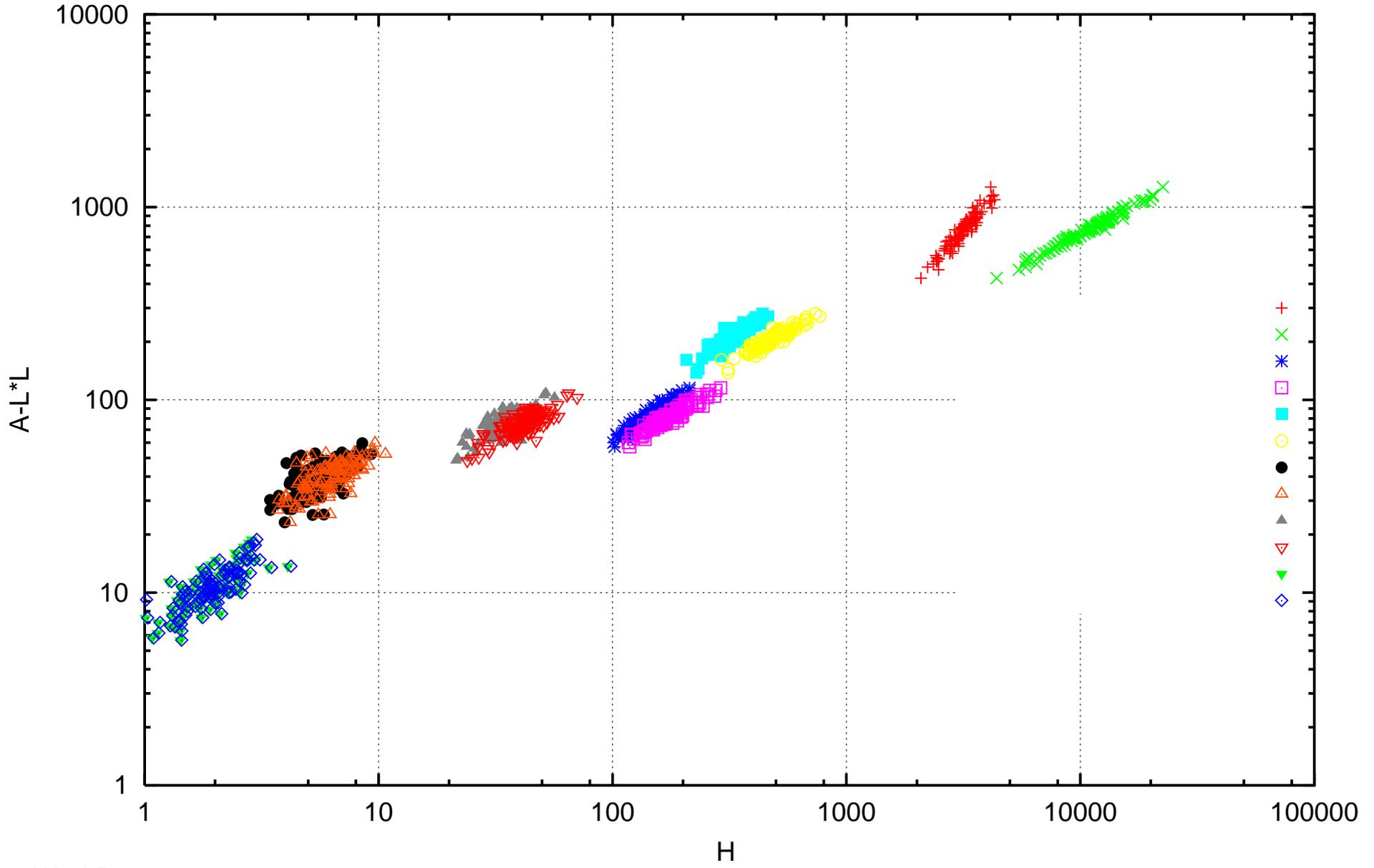
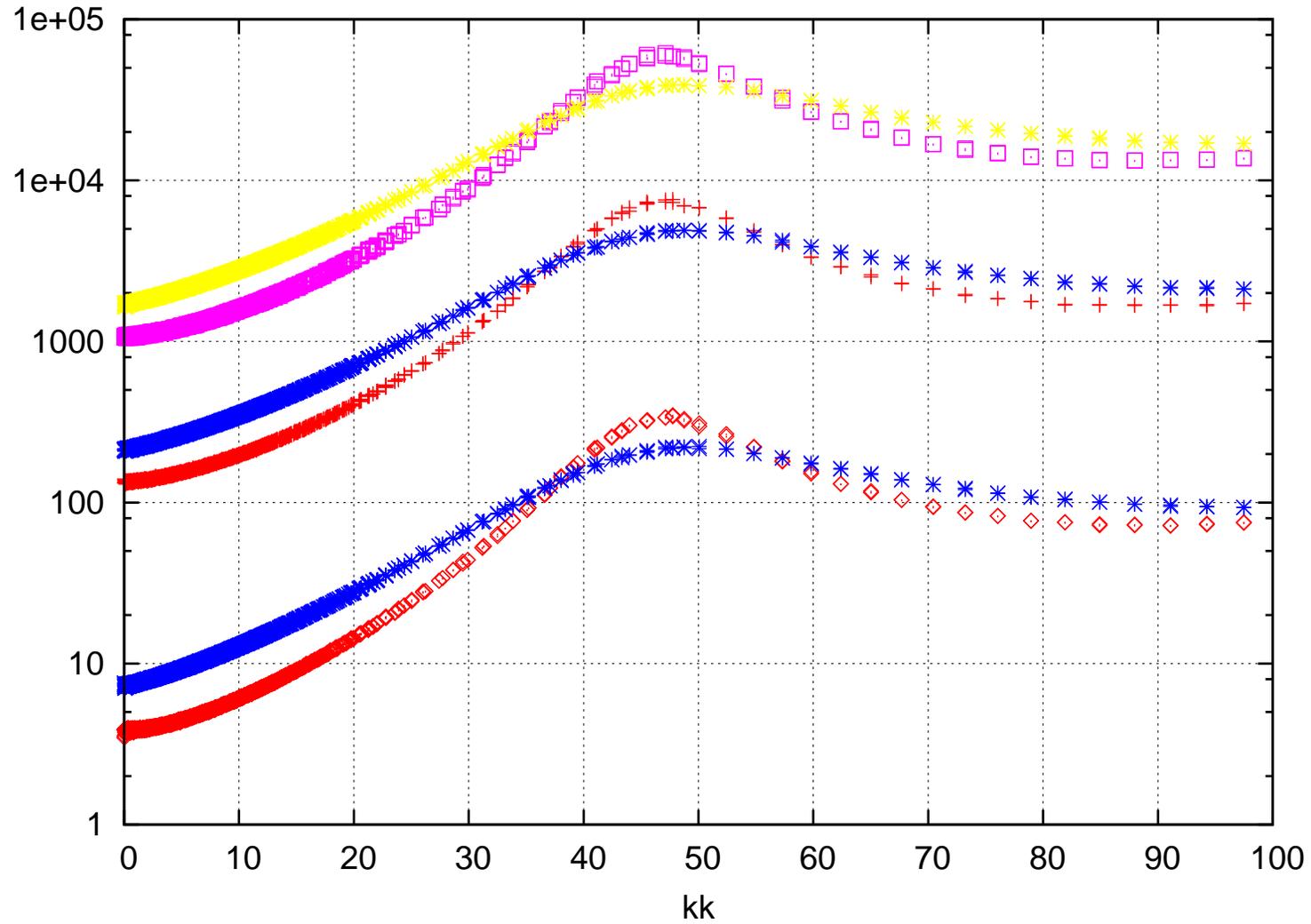
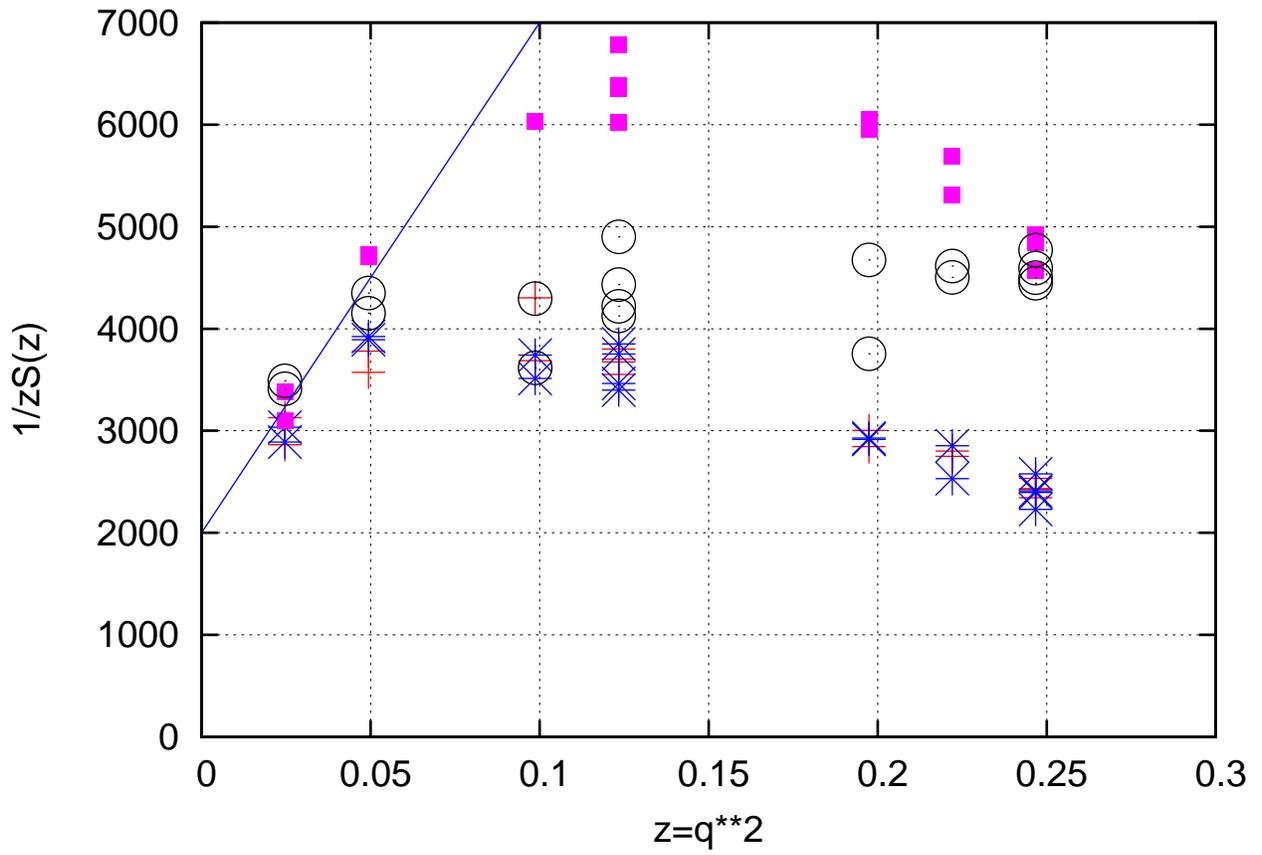
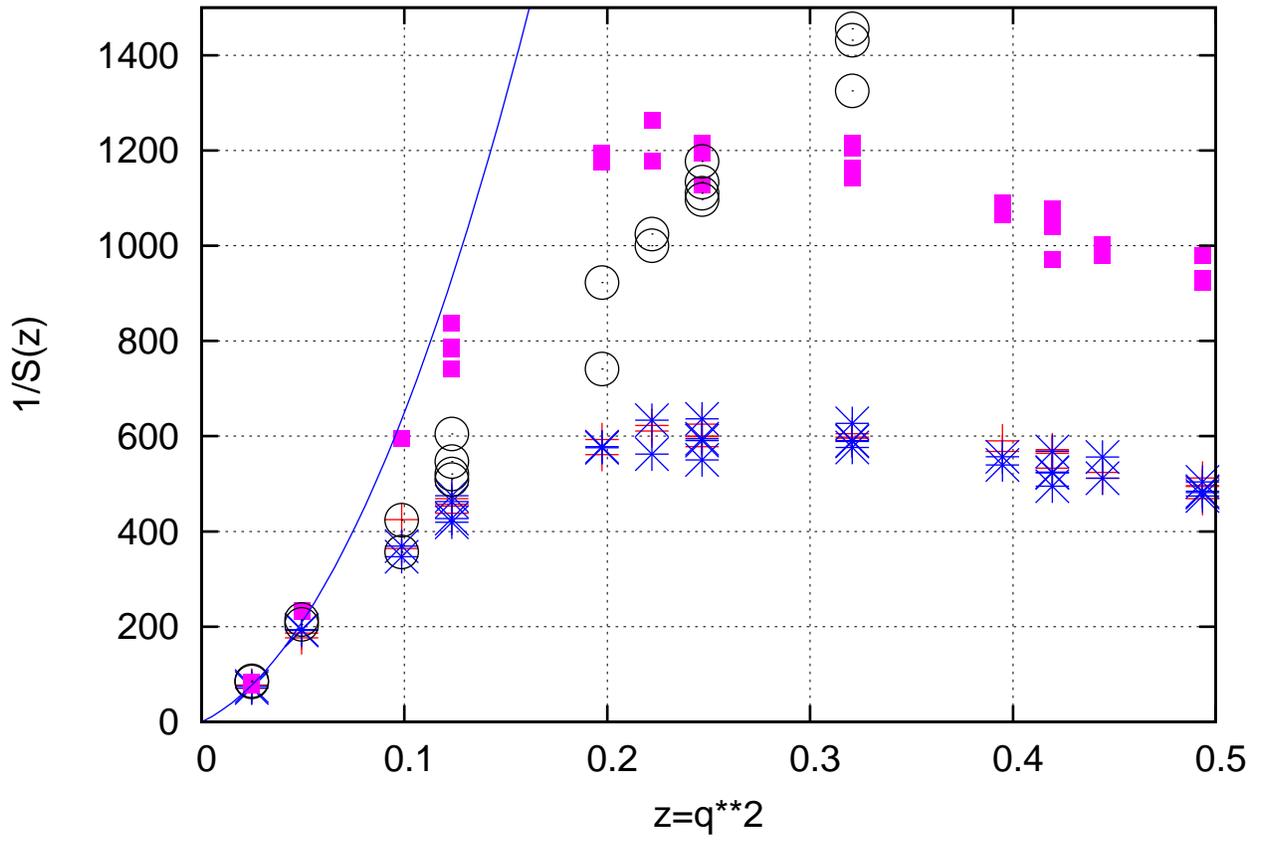
\bye